\documentclass[preprint,nofootinbib]{revtex4}%
\usepackage{amssymb}
\usepackage{amsfonts}
\usepackage{amsmath}
\usepackage{amsmath}
\usepackage{graphicx}
\usepackage[usenames]{color}%
\providecommand{\U}[1]{\protect\rule{.1in}{.1in}}
\providecommand{\U}[1]{\protect\rule{.1in}{.1in}}
\definecolor{blue}{rgb}{0,0,1}

\definecolor{red}{rgb}{1,0,0}

\begin{document}
\title{Rotating and accelerating AdS black holes in Einstein-Gauss-Bonnet Gravity}
\author{Andres Anabalon$^1$, Dumitru Astefanesei$^2$, Adolfo Cisterna$^{3,4}$, Fernando Izaurieta$^5$, Julio Oliva$^1$, Constanza Quijada$^1$, Cristian Quinzacara$^5$}

\affiliation{$^1$Departamento de F\'{\i}sica, Universidad de Concepci\'{o}n, Casilla, 160-C,
Concepci\'{o}n, Chile.}

\affiliation{$^2$Pontificia Universidad Católica de Valparaíso, Instituto de Física, Av. Brasil 2950, Valparaíso, Chile}

\affiliation{$^3$Sede Esmeralda, Universidad de Tarapacá, Avenida Luis Emilio Recabarren 2477, Iquique, Chile}

\affiliation{$^4$Institute of Theoretical Physics, Faculty of Mathematics and Physics, Charles University, V Holesovickach 2, 18000 Prague 8, Czech Republic}

\affiliation{$^5$Departamento de Ciencias Exactas, Facultad de Ingeniería, Arquitectura y Diseño, Universidad San Sebastián, Concepción,Chile.}

\begin{abstract}
We present new, exact, rotating and accelerating solutions within the framework of five-dimensional Einstein-Gauss-Bonnet theory at
the Chern-Simons point. The rotating solutions describe black holes characterized by a single
rotation parameter, a mass parameter, and two extra integration constants that can be interpreted as hairs of a gravitational nature. It is noteworthy that the
rotation cannot be removed by a large diffeomorphism. We also extend our procedure to dimension seven, providing a new rotating solution with two rotation parameters pertaining to cubic Lovelock theory with a unique vacuum.
\end{abstract}
\maketitle

\section{Introduction}

Constructing exact, analytic, rotating black hole solutions in the presence of higher curvature terms in higher dimensions has been a challenging endeavor, even within the simple framework of Einstein-Gauss-Bonnet (EGB) gravity. In EGB gravity, the Einstein-Hilbert term is augmented by a quadratic curvature term, constructed in such a manner that the second-order nature of the field equations is maintained \cite{Lovelock:1971yv} (also see \cite{Garraffo:2008hu}). Thus far, within this framework, the only known rotating solutions have been obtained employing a large diffeomorphism via a boost of a planar black hole (see, for instance, \cite{Dehghani:2002wn}, \cite{Dehghani:2003ea}, \cite{Dehghani:2006dh}, \cite{Dehghani:2006cu}), expanding up to first order in the rotation parameter \cite{Kim:2007iw}\footnote{For a four-dimensional rotating and accelerating black hole solution in the low energy limit of string theory with $\mathcal{R}^2$ corrections, accurate up to $\mathcal{O}\left(\alpha^2\right),\mathcal{O}\left(a^3\right), \mathcal{O}\left(\alpha a^2\right)$, refer to \cite{Agurto-Sepulveda:2022vvf}.}, or resorting to numerical methods \cite{Brihaye:2008kh}, \cite{Brihaye:2013vsa}.

One of the challenges arises from a no-go result identified in \cite{Anabalon:2009kq}, indicating that the Kerr-Schild ansatz, which proves fruitful in General Relativity (GR), fails to yield a rotating solution for arbitrary, finite values of the Gauss-Bonnet coupling parameter $\alpha$. However, the referenced work also demonstrates that for a particular value of the Gauss-Bonnet coupling, satisfying $\alpha\Lambda=-\frac{3}{4}$ (where $\Lambda$ represents the cosmological term in the action), the Kerr-Schild ansatz indeed supports a rotating spacetime characterized by two rotation parameters. Notably, this spacetime configuration is non-circular \cite{Anabalon:2010ns}, and within the original coordinate patch in which the solution was derived \cite{Anabalon:2009kq}, it lacks a Killing horizon. Additional investigations into the Kerr-Schild ansatz within the framework of Lovelock gravity are provided in \cite{Ett:2011fy}.
The significance of the relation $\alpha\Lambda=-\frac{3}{4}$ extends to various aspects: in five dimensions with a negative cosmological constant, the gravitational theory can be reformulated as a gauge theory for $SO(4,2)$, featuring a Chern-Simons action for the gauge connection $A$, which accommodates both the vielbein and the spin connection \cite{Chamseddine:1989nu}. Furthermore, under this condition, the theory overcomes causality constraints on shock wave propagation, which otherwise necessitate the inclusion of higher-spin fields \cite{Camanho:2014apa}. Moreover, it admits a unique maximally symmetric solution accommodating dimensionally continued, static black holes \cite{Banados:1993ur}, vacuum wormholes \cite{Dotti:2006cp}, \cite{Dotti:2007az}, and supersymmetric black holes (cf. \cite{GiriMisk} and references therein). Consequently, exploring exact, stationary solutions within this particular parameter regime becomes a natural pursuit.
The objective of the present paper is to introduce the first exact rotating black hole solution within this framework, complemented by a novel accelerating solution.

The paper is structured as follows: Section 2 begins by introducing the theory along with the novel rotating solutions. Causal structure analysis is conducted for various integration constant values, accompanied by the presentation of the effective potential governing radial geodesic motion. In Section 3, we unveil the newly derived accelerating solution within the theory. Notably, physically plausible scenarios can be readily derived from the foundational three-dimensional metric with constant Ricci scalar curvature, as previously established in \cite{Cisterna:2023qhh}. In Section 4, a novel solution is unveiled in seven dimensions, within the cubic Lovelock theory at the Chern-Simons point. This solution is distinguished by two angular momenta and two integration functions. Finally, Section 5 encapsulates our concluding remarks.

\section{The theory and the rotating black hole}

The gravitational theory is described by the action
\begin{equation}\label{action}
I[g] = \int d^{5}x \sqrt{-g} \left( R - 2\Lambda + \alpha \left( R^{2} - 4R_{ab}R^{ab} + R_{abcd}R^{abcd} \right) \right) \ .
\end{equation}
When the cosmological constant is negative, it proves advantageous to express the Chern-Simons relation $\alpha\Lambda=-\frac{3}{4}$ in terms of the AdS radius $l$ associated with the unique, maximally symmetric solution. This relation can be parameterised as follows
\begin{equation}\label{cspoint}
\alpha = \frac{l^{2}}{4} \quad \text{and} \quad \Lambda = -\frac{3}{l^{2}} \ .
\end{equation}
The new rotating solution is determined by the expression
\begin{align}\label{metricarot}
ds^{2} & = l^{2} \cosh^{2}(\rho) \left[ -\frac{(1+r^{2})}{\Xi_{a}}d\tau^{2} + \frac{r^{2}dr^{2}}{(1+r^{2})(r^{2}+a^{2})} + \frac{(r^{2}+a^{2})}{\Xi_{a}}d\phi^{2} + \right. \nonumber\\
& \left. U(r)\left( \frac{d\tau - ad\phi}{\Xi_{a}} + \frac{r^{2}dr}{(1+r^{2})(r^{2}+a^{2})} \right)^{2} \right] + l^{2}d\rho^{2} + l^{2}\cosh^{2}(\rho-\rho_{0})dz^{2} \ ,
\end{align}
where $\Xi_{a}=1-a^{2}$ and
\begin{equation}
U(r) = \mu + \frac{b}{r} \ .
\end{equation}
In the provided context, $\mu$ represents a mass parameter, and $a$ stands as the sole rotation parameter, while $\rho_{0}$ and $b$ denote integration constants interpreted as gravitational hair. These parameters exert influence on the trajectories of test particles navigating the geometry. The coordinate ranges are defined as $-\infty<t<\infty$, $-\infty<\rho<\infty$, $-\infty<z<\infty$, and $0\leq\phi<2\pi$, with periodic identification. Notably, the coordinate $z$ permits trivial compactification since $g_{zz}$ in Equation \eqref{metricarot} never achieves a zero value. The discussion concerning the range of the coordinate $r$ is postponed for now. We restrict ourselves to consider $a^2<1$.
The surfaces characterized by constant $\rho$ and constant $z$ coordinates define Kerr-Schild geometries, which facilitate the integration of field equations. However, it is noteworthy that the Kerr-Schild geometry herein is three-dimensional, as opposed to the typical five-dimensional counterpart, as a means to circumvent the constraints elucidated in \cite{Anabalon:2009kq}.

To analyze the causal structure effectively, it proves beneficial to implement the following change of coordinates
\begin{equation}
\tau = t + a\psi + \int F_{1}(r) dr \quad \text{and} \quad \phi = \psi + at + \int F_{2}(r) dr \ ,
\end{equation}
where
\begin{equation}
F_{1}(r) = \frac{(\mu r + b)r}{(r^{2} + 1)(r^{4} + (a^{2} - \mu + 1)r^{2} - br + a^{2})} \quad \text{and} \quad F_{2}(r) = a\frac{(r^{2} + 1)}{(r^{2} + a^{2})}F_{1}(r) \ .
\end{equation}
In these transformed coordinates, the metric \eqref{metricarot} assumes the form
\begin{equation}
ds^{2} = l^{2}\cosh^{2}(\rho) \left[ -N^{2}dt^{2} + \frac{dr^{2}}{N^{2}} + r^{2}\left(d\psi + N^{\psi}dt\right)^{2}\right] + l^{2}d\rho^{2} + l^{2}\cosh^{2}(\rho - \rho_{0})dz^{2} \label{BLmetric}\ ,
\end{equation}
where
\begin{equation}
N^{2} = r^{2} - M - \frac{b}{r} + \frac{j^{2}}{4r^{2}} \quad \text{and} \quad N^{\psi} = -\frac{j}{2r^{2}} \ ,
\end{equation}
with $M = \mu - 1 - a^2$ and $j = 2a$. Notably, these coordinates assume the Boyer-Lindquist form, as the Killing vectors $\partial_{t}$ and $\partial_{\psi}$ are orthogonal to the non-Killing ones, thereby ensuring the spacetime's circularity. When $b = 0$, the spacetimes characterized by constant $\rho$ and constant $z$ correspond to three-dimensional BTZ black holes \cite{Banados:1992wn}, featuring event ($r_+$) and Cauchy ($r_-$) horizons located at
\begin{equation}\label{horbtz}
r_{\pm}^{\text{BTZ}} = \sqrt{\frac{M \pm \sqrt{M^2 - a^2}}{2}} \ .
\end{equation}
It is well-established that in this scenario, the rotation can be removed via a large diffeomorphism (boost) on the $(t,\psi)$ plane. Remarkably, when $b$ does not equal zero, it can be demonstrated that a general linear transformation of the form $t \rightarrow \alpha t + \beta\psi$ and $\psi \rightarrow \gamma t + \sigma\psi$ yields a static spacetime (where the $dtd\psi$ term vanishes) only if
\begin{equation}
(\alpha\beta - \gamma\sigma)r^{3} + (-M\alpha\beta + \alpha\sigma a + \beta\gamma)r + \alpha\beta b = 0 \ ,
\end{equation}
which is impossible for generic values of the radial coordinate $r$ when $b$ is non-zero. Consequently, our newly derived solution inherently possesses rotation that cannot be nullified via a large boost on the $(t,\psi)$ plane.
The spacetime described by Equation \eqref{BLmetric} exhibits a curvature singularity at $r=0$, as evidenced by the computation of the Kretschmann invariant as $r\rightarrow0$
\begin{equation}
R_{abcd}R^{abcd} = \frac{6b^{2}}{l^{4}\cosh^{2}\rho}\frac{1}{r^{6}} + \mathcal{O}(1) \ .
\end{equation}
As $\rho\rightarrow\pm\infty$, the spacetime asymptotically approaches locally AdS$_{5}$. This configuration can be understood as a non-homogeneous string, with the $\rho$ coordinate delineating the string, foliated by $\rho=$ constant spacetimes representing rotating black holes. This interpretation arises from the potential existence of up to two zeros of the function $g^{rr}$, denoted as $r_{+}$ and $r_{-}$, which correspond to event and Cauchy horizons, respectively. When $b=0$, these horizons lie at the positions indicated in Equation \eqref{horbtz}. However, with a non-vanishing parameter $b$, their positions are altered, leading to the presence of horizons even when $M<|j|$. While the horizons locations can be expressed algebraically, it proves more instructive to examine the family of plots depicted in Figure 1.
\begin{figure}[h]
\centering
\includegraphics[width=3.0in]{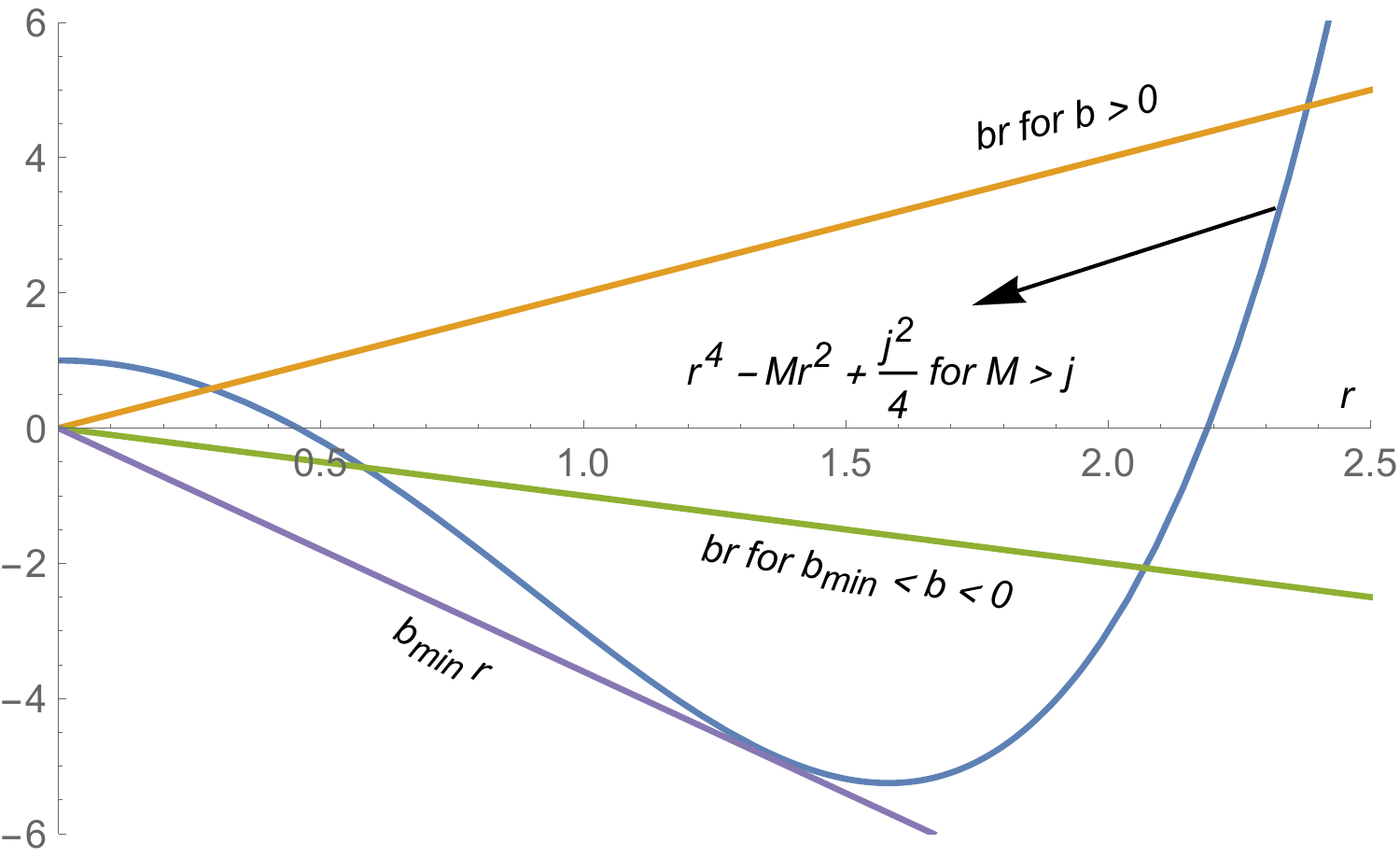} \quad
\includegraphics[width=3.0in]{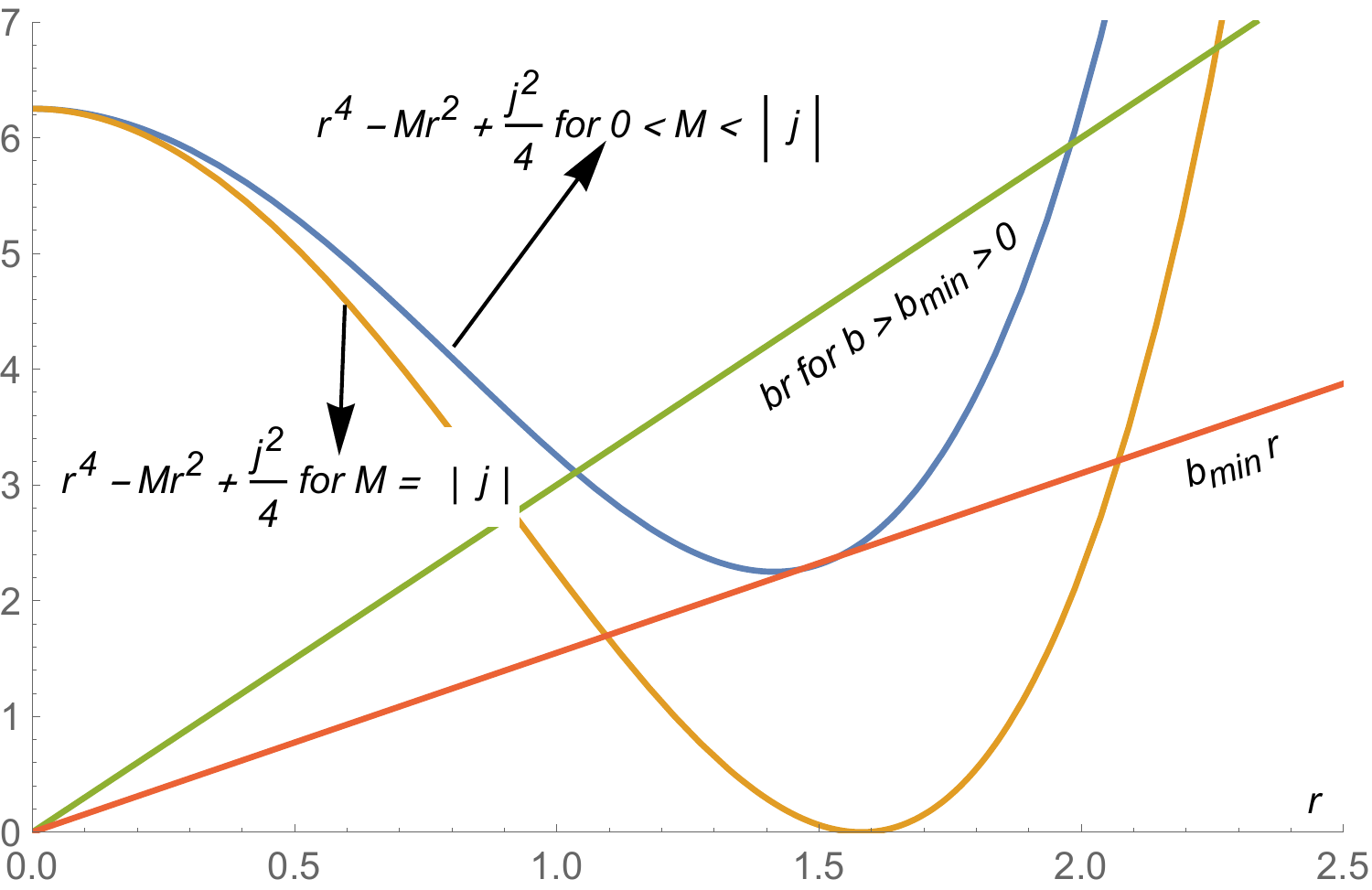} \qquad
\includegraphics[width=3.0in]{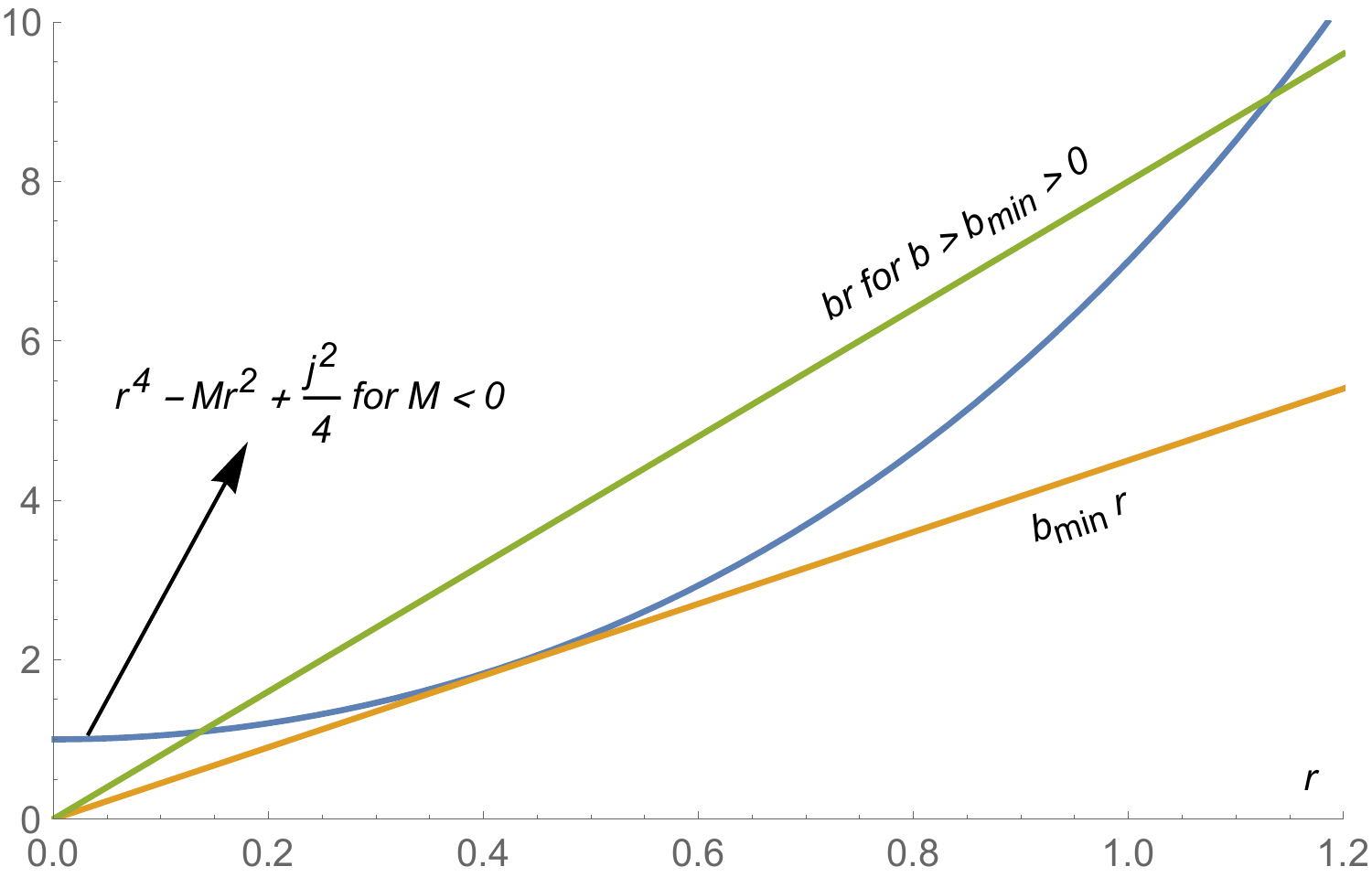}
\caption{Different cases for the functions $r^4-Mr^2+j^2/4$ and $br$ for diverse values of the parameters, versus the radial coordinate $r$. The intersections of the straight lines ($br$) and the quartics shows the existence of a horizon, located at the value of the horizontal coordinate of the intersection point. Notice that since the quartic dominates for large $r$, the three-dimensional spacetime spanned by the coordinates $(t,r,\phi)$ is always asymptotically AdS$_3$, the largest horizon is an event horizon and the smallest, when it exists, it is a Cauchy horizon (see bulk of the manuscript for detailed explanations).}
\label{Figure01}
\end{figure}

Figure 1 illustrates plots of the quartic function $r^4 - Mr^2 + j^2/4$ alongside straight lines of the form $br$, versus the coordinate $r$, showcasing different parameter values. A horizon emerges each time the straight line intersects the quartic, as these intersections correspond to zeroes of the $g^{rr}$ component of the metric \eqref{BLmetric}. The outer horizon consistently manifests as an event horizon, while the inner horizon serves as a Cauchy horizon. This determination arises from the dominance of the quartic for large $r$, rendering $\partial_t$ timelike in the asymptotic region as $r\rightarrow\infty$.
The upper-left panel in Figure 1 delineates the various possible horizons for $M > |j|$ across diverse $b$ values. Notably, there exists a minimum $b=b_{\text{min}}$ value, below which an extremal black hole emerges, while above it, both an event and a Cauchy horizon materialize for any $b > b_{\text{min}}$. When $b$ falls below $b_{\text{min}}$, the spacetime describes a naked singularity.
In the upper-right panel of Figure 1, distinct curves are depicted for scenarios where $M = |j|$ and $0 < M < |j|$. For zero gravitational hair $b$, the spacetime embodies an extremal black hole in the former case, while it presents a naked singularity in the latter. The introduction of a non-zero, positive $b > b_{\text{min}} > 0$ parameter yields an extremal horizon for $M < |j|$. Furthermore, for larger values of the minimum $b > 0$ in this scenario, the spacetime once again develops both an event and a Cauchy horizon.
Finally, the lower panel of Figure 1 illustrates a scenario akin to the upper-right panel, albeit for $M < 0$.

When an event horizon exists, it is possible to demonstrate the presence of an ergoregion surrounding it, with its outer boundary located at $r=r_e$, where
\begin{equation}
\eta(r_e) = r_e^2 - M - \frac{b}{r_e} = 0 \ .
\end{equation}
The horizon generator is denoted as
\begin{equation}
\xi = \partial_t + \Omega_h\partial_\phi \ ,
\end{equation}
where the angular velocity of the horizon is given by $\Omega_h = \frac{j}{2r_+^2}$ and $r=r_+$ signifies the event horizon's location. Utilizing this information, one can compute the black hole's temperature
\begin{equation}
T = \frac{\kappa}{2\pi} = \frac{4r_+^4 + 2br_+ - j^2}{8\pi r_+^3} \ .
\end{equation}
Here, $M$ has been replaced as a function of $(r_+, b, j)$. As anticipated, increasing the angular momentum value diminishes the black hole's temperature, as it approaches extremality.

Before delving into the accelerated solution, let us analyze the timelike geodesic orbits of test particles on the $\rho=0$, $z=0$ submanifold. The Killing vectors $T=\partial_t$ and $\Phi=\partial_\psi$ yield conserved quantities along the geodesic motion, denoted as $E=-T^\mu U_\mu$ and $L=\Phi^\mu U_\mu$, where $U^\mu$ represents the affinely parameterized velocity vector $U^\mu=\left(\dot{t},\dot{r},\dot{\phi},\dot{\rho},\dot{x}\right)$. Utilizing these conserved quantities alongside the normalization condition $U^\mu U_\mu=-1$ leads, as customary, to an equation of the form
\begin{equation}
g_{rr}\dot{r}^2=U(r):=-1+\frac{(g_{\phi\phi}E^2+L^2g_{tt}+2g_{t\phi}LE)}{g_{t\phi}^2-g_{\phi\phi}g_{tt}} \ ,
\end{equation}
where we define an effective potential, which further simplifies to
\begin{equation}
U(r)=-1+\frac{4((E^2-L^2)r^3+L(-jE+LM)r+bL^2)}{rl^2(4r^4-4Mr^2-4rb+j^2)} \ .
\end{equation}
A comprehensive exploration of the geodesic structure of this spacetime will be provided in forthcoming work.

\section{Accelerated solution}

In the realm of non-static solutions, General Relativity accommodates accelerating black holes \cite{levicivita}-\cite{bonnor}. In asymptotically flat scenarios, such spacetimes feature both event and acceleration (Rindler) horizons. The maximal analytic extension portrays two black holes undergoing acceleration, either propelled apart by a strut positioned in between or pulled by a string from infinity. Apart from their inherent significance, these configurations played a pivotal role in the conceptualization of the five-dimensional Black Ring \cite{Dowker:1993bt,Emparan:2001wn}. Hence, a pertinent inquiry arises: can one construct accelerating black holes within the framework of higher curvature gravity? Notably, such investigations have been conducted in the context of four-dimensional C-metric solutions, along with their conformally coupled scalar field counterparts \cite{AnibalScarlett}.
Given the focus of this manuscript and recognizing the enlarged solution space in Einstein-Gauss-Bonnet theory at the Chern-Simons point compared to cases with generic couplings, we explore the existence of accelerating solutions within the theory described by Equation \eqref{action} at the Chern-Simons point \eqref{cspoint}. We employ an ansatz akin to Equation \eqref{metricarot}, wherein the three-dimensional metric within the square bracket of Equation \eqref{metricarot} is replaced by an accelerating, three-dimensional spacetime. Our task is facilitated by recent findings in \cite{Cisterna:2023qhh}, where an accelerating three-dimensional spacetime was established. This spacetime possesses a constant Ricci scalar but does not exhibit constant Riemann curvature, owing to its sourcing by a conformally coupled scalar field.

Consequently, the new accelerating solution in Einstein-Gauss-Bonnet theory at the Chern-Simons point can be expressed as follows
\begin{align}\label{accelerating}
ds^{2} & =\frac{l^{2}\cosh^{2}\left( \rho\right)}{A^2(x+y)^2}\left(-F(y)dt^2+
\frac{dy^2}{F(y)}+\frac{dx^2}{G(x)}\right) +l^{2}%
d\rho^{2}+l^{2}\cosh^{2}\left( \rho-\rho_{0}\right) dz^{2}
\end{align}
where
\begin{align}
F(y)=&-\sigma(1-y^2)(1-\xi y)+\frac{1}{A^2}\ ,\ G(x)=\sigma (1-x^2)(1+\xi x)\ .
\end{align}
Here, $A$ represents the acceleration parameter and $\xi$ denotes an integration constant. The $\sigma$ parameter, which can take the values $\pm1$, simply denotes, along the lines of \cite{Cisterna:2023qhh}, the family of solutions taken into consideration.
This metric, for $\sigma=-1$, describes a three-dimensional accelerating spacetime with a constant Ricci scalar equal to $-6$, warped along an extended direction parameterized by the coordinate $\rho$. In essence, the constant $(\rho,z)$ sections embody accelerating spacetimes. The potential causal structures can be readily obtained from Ref. \cite{Cisterna:2023qhh}.

\section{Extension of the rotating solution to seven dimensions}

Both the rotating solution \eqref{metricarot} and the accelerating solution \eqref{accelerating} derived within the framework of the Einstein-Gauss-Bonnet (EGB) theory, as presented in the preceding sections, can be interpreted as uplifts to five dimensions of three-dimensional spacetimes characterized by a constant Ricci scalar of $-6$. Building upon the success of this approach, we extend our analysis to seven dimensions by considering the spacetime metric 
\begin{align}\label{lade7}
ds^{2}_7  &  =l^{2}\cosh^{2}\left(  \rho\right) ds^2_5  +l^{2}%
d\rho^{2}+l^{2}\cosh^{2}\left(  \rho-\rho_{0}\right)  dx^{2},
\end{align}
in cubic Lovelock theory with a unique vacuum, of which the field equations can be written in a compact manner as
\begin{equation}\label{cubic}
E_{\mu }^{\nu }=\delta _{\mu \beta _{1}...\beta _{6}}^{\nu \alpha
_{1}...\alpha _{6}}\bar{R}_{\ \ \alpha _{1}\alpha _{2}}^{\beta _{1}\beta
_{2}}\bar{R}_{\ \ \alpha _{3}\alpha _{4}}^{\beta _{3}\beta _{4}}\bar{R}_{\ \
\alpha _{5}\alpha _{6}}^{\beta _{5}\beta _{6}}=0\ .
\end{equation}%
Here
\begin{equation*}
\bar{R}_{\ \ \alpha _{1}\alpha _{2}}^{\beta _{1}\beta _{2}}=R_{\ \ \alpha
_{1}\alpha _{2}}^{\beta _{1}\beta _{2}}+\frac{1}{l^{2}}\delta _{\alpha
_{1}\alpha _{2}}^{\beta _{1}\beta _{2}}\ .
\end{equation*}
This theory can also be formulated as a gauge theory with a fiber structure for the group $SO(6,2)$, where, as in dimension five, the vielbein and the spin connection arise as components of a gauge connection valued on the corresponding Lie algebra \cite{Chamseddine:1989nu}, \cite{Hassaine:2016amq}. It is worth mentioning that in this context, a rotating solution with three equal rotation parameters was already constructed in \cite{Cvetic:2016sow}. As shown below, our setup given by the metric \eqref{lade7}, allows us to capture rotating spacetimes with two different rotation parameters, in dimension seven.

In a similar manner as for EGB case, we consider the five dimensional Kerr-Schild metric $ds_5$ in \eqref{lade7}, with
\vspace{0.2cm}
\begin{align}
ds_5^{2}  &  =-\left(  1+\frac{r^{2}}{L^{2}}\right)  \frac
{\Delta\left(  \mu\right)  }{\Xi_{a}\Xi_{b}}dt^{2}+\frac{r^{2}\rho^{2}dr^{2}%
}{\left(  1+\frac{r^{2}}{L^{2}}\right)  \left(  r^{2}+a^{2}\right)  \left(
r^{2}+b^{2}\right)  }+\frac{\rho^{2}d\mu^{2}}{\Delta\left(  \mu\right)
\left(  1-\mu^{2}\right)  }\nonumber\\
&  +\frac{(r^{2}+a^{2})\left(  1-\mu^{2}\right)  }{\Xi_{a}}d\phi^{2}%
+\frac{(r^{2}+b^{2})\mu^{2}}{\Xi_{b}}d\psi^{2}\nonumber\\
&+F(r,\mu)\left(\frac{\Delta(\mu)}{\Xi_{a}\Xi_{b}}dt+\frac{r^{2}\rho^{2}dr}{\left(
1+\frac{r^{2}}{L^{2}}\right)  \left(  r^{2}+a^{2}\right)  \left(  r^{2}%
+b^{2}\right)  }+\frac{a\left(  1-\mu^{2}\right)  }{\Xi_{a}}d\phi+\frac
{b\mu^{2}}{\Xi_{b}}d\psi\right)^2\ ,
\end{align}
where, as usual $\Xi_{a}=1-\frac{a^{2}}{L^{2}},\ \Xi_{b}=1-\frac{b^{2}}{L^{2}}$, and $\Delta\left(  \mu\right)  =\Xi_{a}\mu^{2}+\Xi_{b}\left(  1-\mu^{2}\right)
$.
The field equations of cubic Lovelock theory, with a unique vacuum \eqref{cubic}, are fulfilled provided the Kerr-Schild function $F(r,\mu)$ satisfies the quadratic polynomial equation
\vspace{0.2cm}
\begin{equation}
\left( \frac{4r^{2}-\Sigma ^{2}}{2\Sigma ^{2}}\right) F\left( r,\mu \right)
^{2}+\frac{3}{L^{2}}\left( L^{2}-1\right) \Sigma ^{2}F\left( r,\mu \right)
=6m\left( \mu \right) +\frac{d\left( \mu \right) }{r}-\frac{3}{L^{2}}\left(
L^{2}-1\right) r^{2}\left( 10\Sigma ^{2}-7r^{2}\right) , 
\end{equation}
where we have defined
\begin{equation}
\Sigma ^{2}=r^{2}+a^{2}\mu ^{2}+b^{2}\left( 1-\mu ^{2}\right) .
\end{equation}
In this case, $L$ is an integration constant, while $m(\mu)$ and $d(\mu)$, are integration functions, which play the role of the constants $m$ and $b$ of EGB solution \eqref{metricarot}. For general values of the integration constants and functions, it was shown in \cite{Anabalon:2010ns} that the five dimensional spacetime $ds^2_5$ is non-circular, a property that is inherited by the seven-dimensional spacetime \eqref{lade7}.

\section{Conclusions}

In this work, we have presented the first exact rotating black hole solution of Einstein-Gauss-Bonnet gravity at the Chern-Simons point. Additionally, we introduced a new accelerating solution and extended the rotating spacetime to cubic Lovelock theory with a unique vacuum. In dimension five, the solution possesses a single rotation parameter, and the geometry can be understood as the uplift of a three-dimensional black hole with constant Ricci scalar to dimension five, through the inclusion of a suitable warp factor. Similarly, the rotating black hole in dimension seven arises as the oxidation of a five-dimensional geometry that satisfies the trace of the EGB field equations with a unique vacuum.

The five-dimensional black hole metric \eqref{BLmetric} is circular and characterized by four integration constants: a mass parameter $M$, an angular momentum parameter $j$, and two gravitational hairs $b$ and $\rho_0$. For non-vanishing $b$, we demonstrated that the crossed term in the metric $g_{t\psi}$ cannot be removed by a boost, resulting in a spacetime with intrinsic rotation. Additionally, we showed that the geodesic equation is separable, obtaining the effective potential for radial motion, which can be utilized for future analyses of particle motion on this rotating geometry. For generic parameter values, the event horizon is surrounded by an ergoregion. Our new solutions satisfy the vacuum field equations of Einstein-Gauss-Bonnet theory at the Chern-Simons point. Similar oxidations of rotating solutions in the presence of non-minimally coupled scalar fields within the Horndeski family, leading to exact rotating solutions in Einstein-Gauss-Bonnet gravity in arbitrary dimensions, were constructed in \cite{rothorn}.

Since the component $g_{t\psi}$ of the black hole metric \eqref{BLmetric} is governed by the integration constant $j$, it is natural to identify such a constant with the angular momentum. In EGB at the Chern-Simons point, a conserved charge can be obtained by applying Noether's theorem to a suitably regularized action principle in asymptotically locally AdS spacetimes \cite{MoraOlea,Kofinas:2006hr}, which can be applied to obtain the energy of the dimensionally continued black holes \cite{Banados:1993ur} as well as to compute the energy content of the wormhole geometries that exist in EGB at the Chern-Simons point \cite{Dotti:2006cp}. Since our new rotating black hole \eqref{BLmetric} possesses two asymptotically locally AdS regions as $\rho\rightarrow\pm \infty$, one may apply the formulae of \cite{MoraOlea,Kofinas:2006hr} to obtain the mass $Q\left(\partial_t\right)$ and the angular momentum $Q\left(\partial_\psi\right)$ as conserved charges associated with the corresponding isometries. However, one can observe that such computations yield infinite values for the conserved charges, with the divergence stemming from the non-compactness of the radial coordinate $r$ as $r\rightarrow\infty$. In this region, the geometry is not asymptotically locally AdS, thus rendering the regularization prescription of \cite{MoraOlea,Kofinas:2006hr} inapplicable. To explore the thermodynamics of our new rotating solutions, it would be valuable to develop a new regularization scheme for Einstein-Gauss-Bonnet gravity at the Chern-Simons point, adapted to the asymptotic behavior observed as $r\rightarrow\infty$, which may yield finite values for the charges.

\section*{Acknowledgments}
We thank Felipe Agurto, Eloy Ayon-Beato, Gaston Giribet, Cristian López and Marcelo Oyarzo for enlightening comments. The work of A.A. is supported in part by the FONDECYT grants 1200986, 1210635, 1221504, 1230853 and 1242043. The work of D.A. is supported in part by the FONDECYT grant 1242043. The work
of A.C. is partially funded by FONDECYT Regular grant No. 1210500 and Primus grant PRIMUS/23/SCI/005 from Charles University. F.I. is supported by FONDECYT grant 1211219 from the Government of Chile. J.O. is partially supported by FONDECYT Grant 1221504. The work of C. Quijada was partially supported by ANID Fellowship 21191868.  C. Quinzacara is supported by FONDECYT grant 11231238.

\end{document}